DOI: 10.1039/C6CE02348G

ARTICLE

# Growing low-dimensional supramolecular crystals directly from 3D particles

Alexander Eberle,[a] Adrian Nosek,[b] Johannes Büttner,[c] Thomas Markert,[d] and Frank Trixler,[a,c]

We show that one-dimensional (1D) nanostructures and two-dimensional (2D) supramolecular crystals of organic semiconductors can be grown on substrates under ambient conditions directly from three-dimensional (3D) organic crystals. The approach does not require dissolving, melting or evaporating of the source crystals and is based on the Organic Solid-Solid Wetting Deposition (OSWD). We exemplify our approach by the pigment quinacridone (QAC). Scanning Tunnelling Microscopy (STM) investigations show that the structures of the resulting 2D crystals are similar to the chain arrangement of the alpha and beta QAC polymorphs and are independent of the 3D source crystal polymorph (gamma). Furthermore, distinct 1D chains can be produced systematically.

## Introduction

One of the most promising area of research, concerning the development of future electronic devices, deals with the so-called carbon-based nanoelectronics.[1-8] In this regard, utilizing the outstanding electronic properties of graphene offers both, an enhanced performance increase and the development of new electronic device types, as flexible or inkjet printed electronics.[1-10] Further, the semiconductive parts of a carbon-based nanocircuitry can be fabricated by organic semiconductor molecules. Such 2D supramolecular arrays induce a bandgap if adhered to the graphene substrate.[6-8] Thus, the assembly of supramolecular architectures consisting of organic semiconductors via 2D crystal engineering is of special interest as far as the construction of functional nanosystems is concerned.[10-15] Functional nanosystems, like nano-scale transistors, are created by the self-assembling of small organic molecules via programmable non-covalent interactions, as hydrogen bonding, Van-der-Waals, π–π stacking, and electrostatics.[10-15]

Today's standard technologies for the bottom-up assembly of supramolecular arrays are mainly based either on vapour deposition[16] or on liquid phase deposition techniques.[17] However, processing organic semiconductors via vapour deposition methods, like the organic molecular beam deposition,[17-18] is often challenging due to their thermal instability during the vacuum sublimation process.[19-21] Also, since most organic pigments with promising semiconductive properties are insoluble in almost all liquid media, liquid phase deposition techniques, like drop-casting or spin-coating,[19] cannot be applied without a chemical functionalization to enable the dissolution of such organic pigments.[19-21] However, the custom synthesis of functionalised semiconductors is costly and expensive, in particular in relation to the purchase prices of the standard pigments, that are already used in the industry. In contrast to the above mentioned limitations, we introduce the Organic Solid-Solid Wetting Deposition (OSWD)[20-22] as a novel, environmental friendly, cheap, and up-scalable technology to assemble 2D supramolecular semiconductive arrays. Compared with the time-consuming and expensive standard manufacturing techniques, based on either vapour- or liquid phase deposition, the OSWD is a very straightforward technique to perform: under ambient conditions, the powdered organic semiconductor is dispersed in a dispersing agent and then drop-casted on a substrate, e.g. graphite, graphene or molybdenum disulphide ($MoS_2$). Immediately afterwards, highly ordered supramolecular arrays covering the substrate surface are formed. The self-assembly is driven by a solid-solid wetting effect,[23-25] with a gradient of the surface free energy acting as the driving force.

In this publication we show that via the OSWD technique, 3D crystalline particles can be transformed directly into several, distinctly defined 1D and 2D supramolecular structures. We exemplify our approach using the organic semiconductor quinacridone (QAC), a cheap and commercially available pigment with promising electrical properties, low toxicity, excellent physical and chemical stability and biocompatibility for applications within the living organism.[26-32] The chemical structure of a single QAC molecule is shown in Fig. 1. The linear QAC molecules built up stable 3D crystal structures by connecting with their neighbours by four hydrogen bonds of the type NH···O=C, forming four possible polymorphs: $α^I$, $α^{II}$, β, and γ.[36] A detailed analysis of the related crystal structures of these polymorphs using X-ray powder diffraction was provided by

[a.] Department für Geo- und Umweltwissenschaften and Center for NanoScience (CeNS), Ludwig-Maximilians-Universität München, Theresienstraße 41, 80333 München, Germany.
[b.] Department of Physics & Astronomy, University of California, Riverside, 900 University Ave., Riverside, CA 92521, USA
[c.] TUM School of Education, Technische Universität München and Deutsches Museum, Museumsinsel 1, 80538 München, Germany
[d.] Institut für Theoretische Chemie, Universität Ulm, Albert-Einstein-Allee 11, 89081 Ulm, Germany

Paulus et al. [36] and by Lincke. [37] The crystal morphology of a 3D QAC crystal, formed by the gamma quinacridone polymorph (γQAC), is shown in Fig. 2, (a). Please note, that below the abbreviation QAC is used for quinacridone in general, usually related to either quinacridone molecules or 1D as well as 2D quinacridone structures, and γQAC is used for the 3D quinacridone polymorph gamma. At standard atmospheric pressure, QAC crystals are entirely insoluble in water,[33-36] as well as in common organic solvents.[28,36] As a result, QAC monolayers could so far only be assembled via organic molecular beam deposition in ultra-high vacuum[32] or by the use of soluble QAC derivates.[28] However, we demonstrate the direct generation of low-dimensional QAC adsorbate crystals from 3D organic crystals via the OSWD technology without a dissolving step and without vacuum sublimation and present a detailed structure analysis of the resulting patterns. In addition, we present results from the OSWD on various substrates, including carbon nanotubes, and show that the dispersing agent can catalyse and direct this crystallisation at the solid-solid interface. Finally, the thermal and temporal stability of the generated low-dimensional structures and the general validity of the presented results shall be discussed.

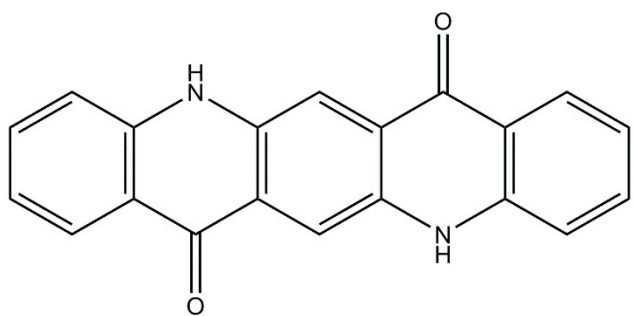

**Fig. 1** Chemical structure of the QAC molecule.

## Results and discussion

**Characterisation of the 1D and 2D crystal structures**

To begin with, we characterise the crystallographic properties of different supramolecular QAC structures that have been grown atop a highly ordered pyrolytic graphite (HOPG) as the substrate, this being done for both the 1D chains as well as for the 2D arrays. The below described structure characterisation is based on empirically determined parameters that have been acquired via the analysis of Scanning Tunnelling Microscopy (STM) scans. Additionally, these parameters and the below presented different types of supramolecular QAC structures have been verified with the aid of force-field calculations, by simulating QAC adsorbate structures atop a graphene substrate. Initially, the QAC molecules are linked in 3D crystals, the molecular arrangement within a γQAC crystal being as shown in the Fig. 2, (b). The OSWD process enables the detachment of molecules from an adsorbed QAC crystal. The detached molecules attach themselves to the substrate surface, where they begin to assemble new supramolecular structures. According to the functional groups of the QAC molecule, QAC builds up supramolecular chains via NH···O=C hydrogen bonds[20] and multiple parallel and side-by-side appearing chains build supramolecular arrays (Fig. 2, (c)). These 1D supramolecular chains and 2D arrays are shown in the simulation in Fig. 2, (c) that is a result of the performed force-field calculations.

Examining how single QAC chains lie side by side to form an array, two possible configurations were observed: the single chains can be located directly next to each other and this condition is termed close-packing QAC chain configuration (Fig. 3, lattice vectors a and b). In addition, there is another case where varying, but definite spacing among the QAC chains is observed – such a configuration being termed as the relaxed QAC chain configuration (Fig. 3, lattice vectors a' and b'). The lattice parameters of this two configurations are:

- Close-packing QAC chain configuration:
  $|a|$ =0,70 ± 0,02 nm; $|b|$ = 1,63 ± 0,02 nm, γ = 87 ± 1°

- Relaxed QAC chain configuration:
  $|a'|$ = $|a|$, $|b'|$ = 2,04 ± 0,02 nm, γ = 88 ± 1°

Investigating the origin of the relaxed QAC chain configuration, we compared the QAC lattice vectors with the graphite lattice vectors ($g_1$ and $g_2$ as in the Fig. 3). For this, we used a STM picture, received directly after scanning the QAC chains (refer Fig. 3, inlay). Converting the lattice vectors of the substrate ($g_1$, $g_2$) into the lattice vectors of the adsorbates (a, b respectively a', b'), we obtained the following transformation matrixes:

$$\begin{pmatrix}a\\b\end{pmatrix} = \begin{pmatrix}10/3 & 4/3\\1 & 7\end{pmatrix}\begin{pmatrix}g_1\\g_2\end{pmatrix} \qquad \begin{pmatrix}a'\\b'\end{pmatrix} = \begin{pmatrix}10/3 & 4/3\\2 & 9\end{pmatrix}\begin{pmatrix}g_1\\g_2\end{pmatrix}$$

The lattice vectors a and a' of the QAC adsorbates are an identic multiple of the graphite vectors and are coincident with the graphite structure, as a result of the strong intermolecular NH···O=C hydrogen bonds of the QAC chain.[20] The vectors b and b' are a whole-number multiple of the graphite vectors and thus are commensurable. The vector b indicates the smallest possible distance between two QAC chains, while the lattice vector b' of the relaxed configuration is the next greater distance, as a consequence of the crystal lattice of the graphite substrate. In the direction of the b and b' vectors, perpendicular to the growth direction of the QAC chains, the Van-der-Waals interactions between the QAC chain and the graphite substrate were found to exceed the Van-der-Waals interactions between neighbouring QAC chains.[20] Greater distances between the QAC chains arise due to missing chains along the periodicity of the b or b' vectors (Fig. 4).

Further investigations revealed, that these different types of QAC chain configurations lead to the formation of three different types of QAC arrays: the arrays with a high packing density (Fig. 5, (a)), the arrays with a low packing density (Fig. 5, (b)) and the lose appearing 1D QAC chains (Fig. 5, (c)). The 2D QAC arrays with high packing density are the result of the close-packing QAC chain configuration. The QAC arrays with low packing density have been found to be the result of either the relaxed QAC chain configuration or the absence of QAC chains. According to Fig. 4 and Fig. 6, QAC wires as well as QAC arrays do not only assemble in a parallel, but in a multi-directional

**Journal Name** ARTICLE

fashion. Analysing the Fourier transformation of various multi-directional arrays, it was further found that QAC domains assemble in six different orientations on the graphite substrate (Fig. 6). Comparing these structure details with the chain arrangements within the QAC crystal-polymorphs, they were found to be independent of their 3D crystalline source, the gamma polymorph. In fact, the via OSWD assembled 2D arrays seem to be similar to the alpha and beta polymorphs of the 3D QAC crystals.[36]

In rare cases, supramolecular QAC bilayer were found, arising as a 2D QAC array is built on top of another one. Further, these two-layer structures were found to occur in two different configurations: a QAC array overgrowing a contiguous QAC array with a different orientation leading to the crisscross structure of two overlapping domains (Fig. 7, (a)). Additionally, the supramolecular QAC chains forming the top layer arising straight above and parallel to the chains of the bottom layer (Fig. 7, (b)), resulting in a plane-parallel bilayer configuration. However, as the top layer can be easily removed using the tip of the STM, the binding energy between the top and the bottom QAC layer is supposed to be considerably lower than the binding energy between the bottom layer and the HOPG surface. Thus we propose, that as the bilayers seem to be instable and prone to mechanical stress, the STM measurement process might destroy these structures before they can be detected. So far, though we've been able to verify the occurrence of these two-layer structures using various sample preparation methods, we still, however, were not successful in directing their growth.

At this point, we'd like to point out once more, that the OSWD is based on a solid-solid wetting effect, with a gradient of the surface free energy being the driving force. Referring to previous publications, we have demonstrated in detail, that there are no detectable dissociated semiconductor-molecules present in the liquid phase of the dispersion in use and thus the OSWD is not related to a solubility phenomenon.[20-21] Additionally, we pointed out that a 2D self-assembly from a liquid phase results in periodic structures, but the observed supramolecular structures arrange themselves in well-defined arrays, displaying varying orientations. Furthermore, if molecules are removed from an adsorbate layer and the liquid phase is still present, then the gap within the periodic structures would immediately be healed by nearby dissociated molecules. However, if a nanoscale gap is induced in via an OSWD self-assembled array, such that the considerably greater semiconductor particles within the liquid phase cannot contact the substrate surface, then this gap remains stable and detectable for days. Also, we have demonstrated, that the time-dependent gap-stability decreases with an increasing gap size, as a result of the increasing probability, that a nearby semiconductive particle fits into the gap.

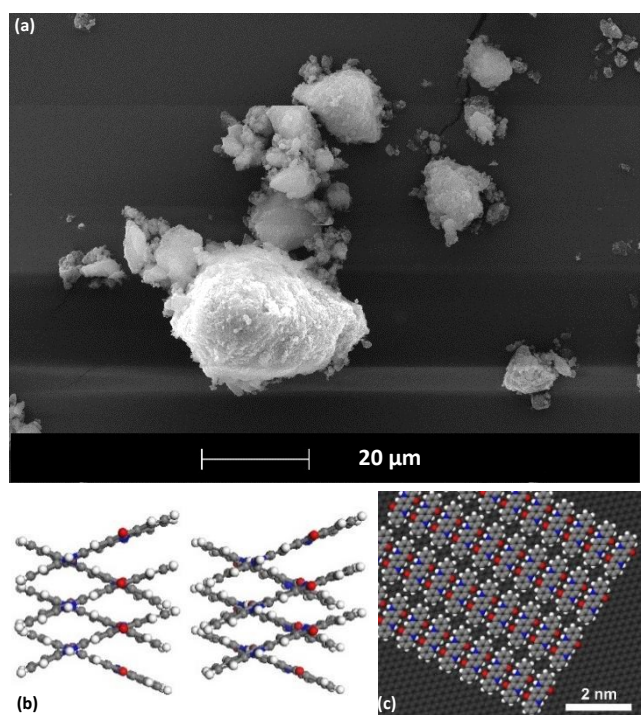

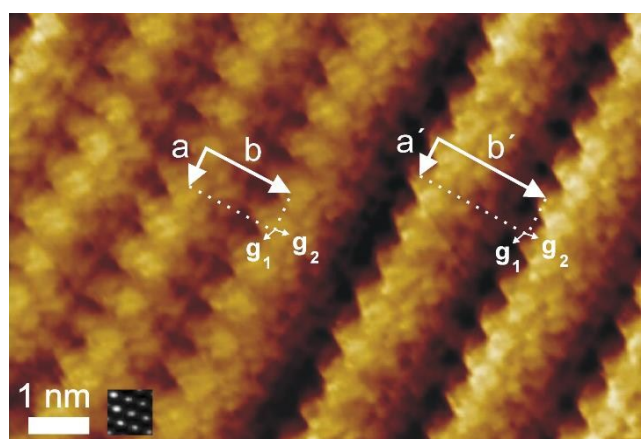

**Fig. 3** STM picture of supramolecular QAC chains on a HOPG displaying variable interspaces. The lattice vectors a and b indicate the unit cell of the close-packing chain configuration of the QAC adsorbate whereas a' and b' indicate the lattice vectors of the relaxed QAC chain configuration. The lattice vectors $g_1$ and $g_2$ of the graphite substrate are added in the correct relation to the vectors of the adsorbate lattice. Inlay: auto-correlated STM image of the HOPG substrate, received directly after scanning the QAC chains.

**Fig. 2** (a) Scanning electron microscopy picture, showing the 3D crystal morphology of γQAC. (b) Model of how the QAC molecules are arranged in the γQAC polymorph. (c) Force-field calculated supramolecular structure, showing how the QAC molecules arrange themselves in 1D supramolecular chains and 2D arrays.

**Journal Name**

ARTICLE

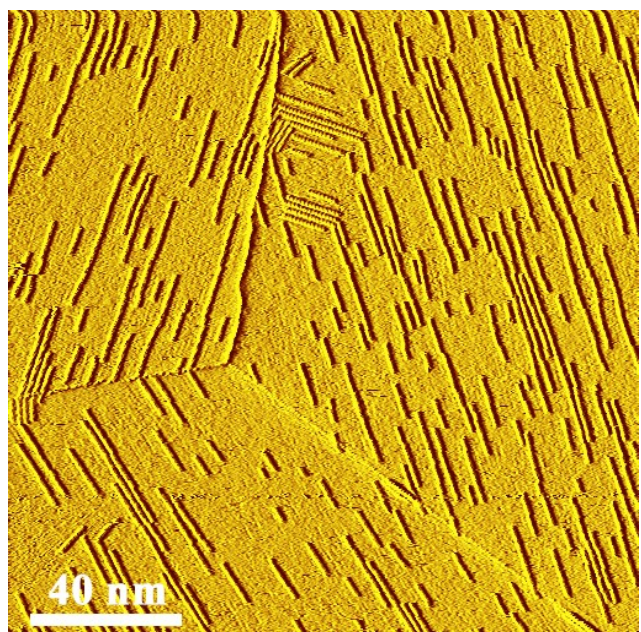

**Fig. 4** STM picture of a HOPG surface that predominantly is covered by 1D QAC chains with varying gaps between them. The average coverage of the whole treated HOPG surface is 24 ± 4 %. Ethylbenzene was used as the catalysing dispersing agent.

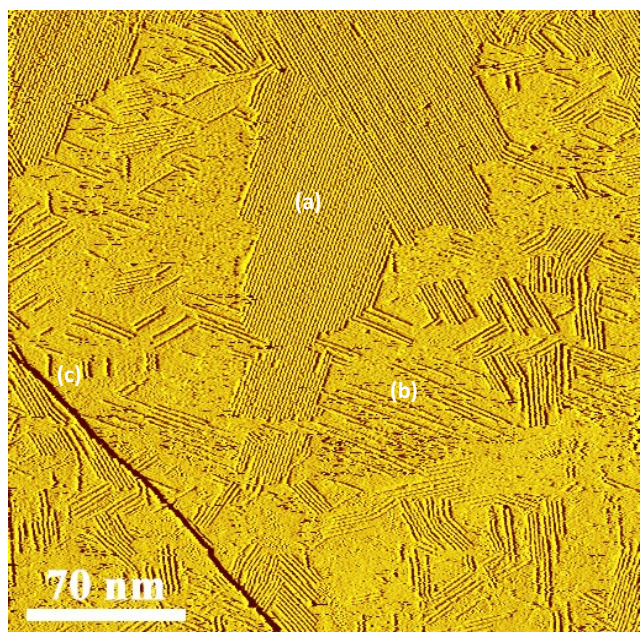

**Fig. 5** STM picture of QAC arrays on a HOPG surface. The dispersing agent anisole was used to catalyse the OSWD, leading to an overall coverage of the surface by QAC arrays of 61 ± 9 % (a) QAC array with a high packing density. (b) QAC array with a low packing density. (c) 1D QAC chain.

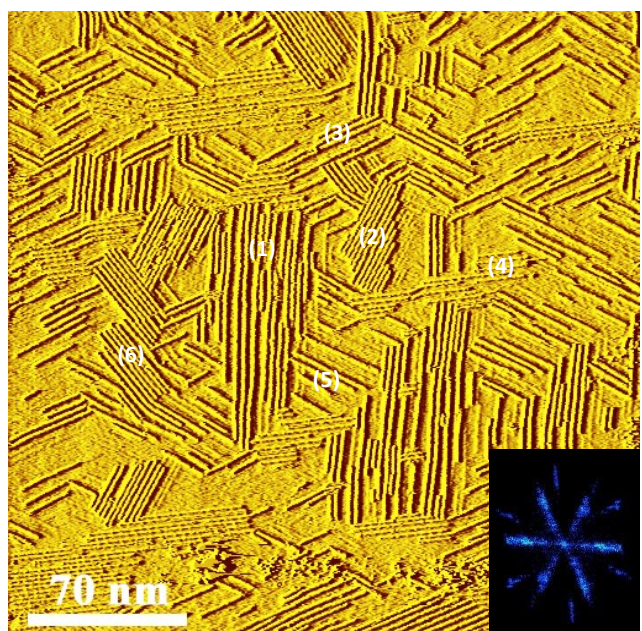

**Fig. 6** The QAC arrays shown in this STM picture have been assembled with the aid of the dispersing agent glycerol. The arrays occur with six different orientations that are shown in (1) – (6). The small picture in the bottom right corner shows the Fast Fourier Transformation of this STM picture.

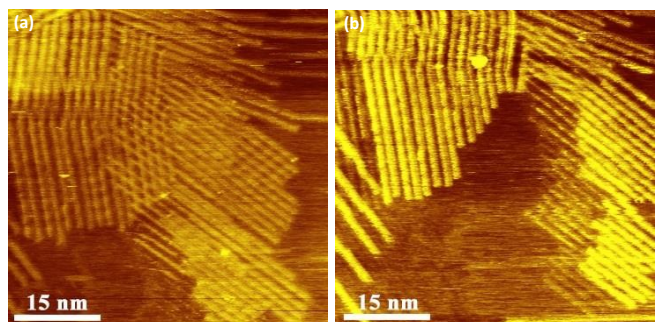

**Fig. 7** (a): Supramolecular QAC bilayer in a crisscross structure of two overlapping QAC arrays. (b): Plane-parallel QAC bilayer, with the QAC chains forming the top layer being on top of and parallel to the chains of the bottom layer.

# Journal Name

## ARTICLE

**Applying the OSWD to alternative substrates: CNTs and MoS$_2$**

The question arose whether the OSWD is capable of generating supramolecular structures on different kinds of substrates as well. Thus, we decided to explore, whether the 1D and the 2D supramolecular QAC configurations that have been discovered so far using a HOPG substrate, can also be detected on other substrates. For this, we began a series of tests using MoS$_2$ as the underlying substrate. The results of our investigations are illustrated in Fig. 8. As can be seen, the single QAC molecules arrange themselves again in 1D chains, with a likewise lattice vector |a| of 0,70 ± 0,02 nm. Furthermore, in analogy to our investigations using HOPG's, multiple parallel and side-by-side appearing chains form arrays displaying various orientations, however, the distance between the single chains (vector b) being now constantly 2,15 ± 0,02 nm (Fig. 8, (a) and (b)). Fast Fourier Transformations further revealed six main array orientations, that are likewise related to the substrate's crystal lattice. An occasionally occurring seventh orientation, as depicted in Fig. 8, (c), was as well revealed, potentially being the result of an array shift along a grain boundary of the substrate. In an alternative approach, we tried to cover multi-walled carbon nanotubes (CNTs). We expected, that as the chemical structure of the CNTs is identical to the one of a HOPG, potentially occurring supramolecular QAC structures on top of them should also display similar configurations. Additionally, we wanted to verify, whether the substrate surface has to be flat or if the OSWD is also applicable to nanoscale curved surface structures. As shown in Fig. 9, the OSWD actually was found to cover the CNTs with supramolecular QAC structures. Exploring the crystallographic properties of these structures revealed, that the 1D QAC chains arrange themselves primarily in the close-packing QAC chain configuration. Thus, the CNTs are closely covered by a QAC monolayer. Additionally, this result underlines the high mechanical stability of the 1D and 2D QAC structures.

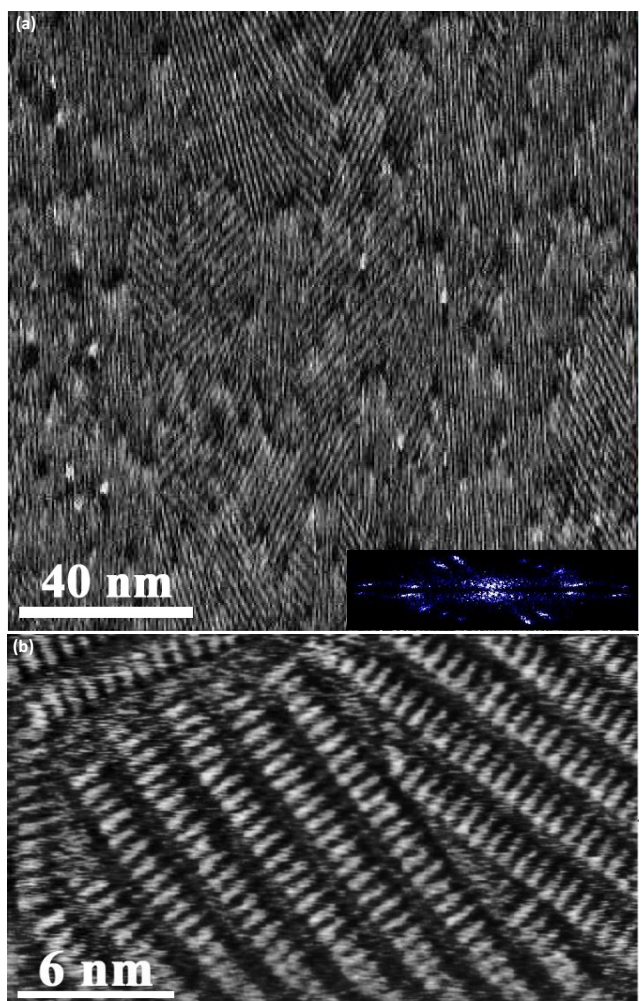

**Fig. 8** (a): Overview STM scan of 2D supramolecular QAC arrays on a MoS$_2$ substrate. The small picture in the bottom right corner shows the related Fast Fourier Transformation. (b) Straightened close-up view of the overview scan, showing that the single QAC molecules arrange themselves in 1D chains. The distance between parallel QAC chains is constant. The dispersing agent in use catalysing the OSWD was octylcyanobiphenyl (8CB).

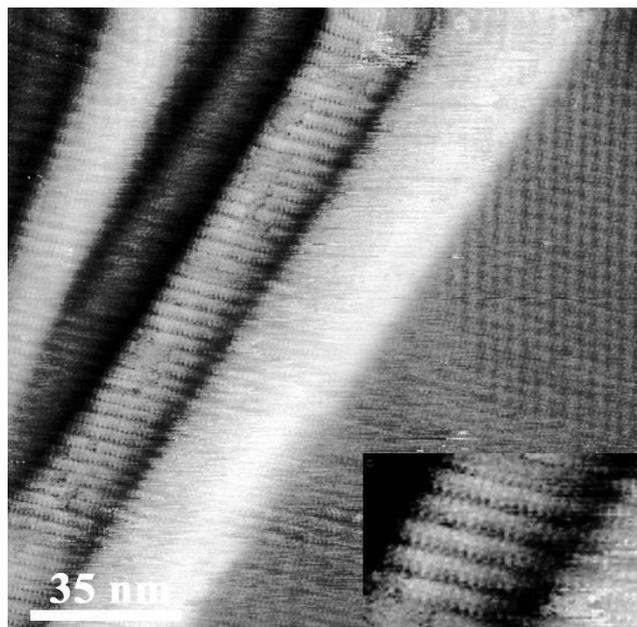

**Fig. 9** STM picture of multi-walled carbon nanotubes (CNTs) that are covered by supramolecular QAC structures. The close-up view in the bottom right corner illustrates, that to cover the CNTs the 1D QAC chains arrange themselves primarily in the close-packing QAC chain configuration.

**The dispersing agent determines structure and quantity of the QAC arrays**

After the determination of the three large-scale configurations of the supramolecular QAC structures (high packing density, low packing density and lose appearing 1D chains), we faced the question, whether these structures occur randomly or if the probability of their occurrence is dependent on the environmental conditions.

First of all, we tried to verify if an OSWD takes place without a dispersing agent. Subject to the current model, the OSWD depends on a gradient of the surface free energy. [20-25] Hence, assuming the situation of a γQAC crystal contacting a HOPG substrate, if the attractive forces towards the substrate surface exceed the binding forces within the adsorbed semiconductor crystal, then an OSWD should take place. Comparing the surface free energies of graphite (54.8 mJ/m$^2$,[43]) and γQAC (49.1 mJ/m$^2$,[44]) revealed, that the binding forces within the graphite surface are stronger than within a γQAC crystal. This fact implies, that inducing an OSWD in ambient conditions could be possible without the necessity to use a dispersing agent. Thus, we linked the powdered γQAC and the HOPG substrate without a dispersing agent in a couple of approaches. However, for all of the cases tried, no QAC arrays could be detected via STM measurements.

Thereupon, we started a series of tests using several polar and non-polar dispersing agents with varying properties. It was observed, that the selection of the dispersing agent had a great influence on the structure, size, and coverage of the assembled QAC arrays. In this regard, propylene carbonate was found to predominantly catalyse the formation of arrays with high packing density and with an achieved coverage of 67 ± 19 % it is the most powerful dispersing agent to catalyse the OSWD process detected so far (Fig. 10). Further, the individual arrays were observed to have small dimensions, as compared to the arrays when assembled with the aid of anisole (Fig. 5). However, anisole was found not to preferentially catalyse a specific array configuration and hence arrays with a high packing density, arrays with a low packing density, and lose appearing 1D QAC chains were observed having approximately the same probability of occurrence. Using anisole, the achieved coverage rate of 61 ± 9 % is still high, but lower than observed in the case of the propylene carbonate sample.

Surprisingly, we were able to trigger the OSWD process by simply using purified water as the dispersing agent. This resulted in the building up of arrays, chiefly with a low packing density, as can be seen in the Fig. 10. In contrast to the polycarbonate and anisole samples, the supramolecular QAC structures built up using purified water were seen covering a considerably lower area of 24 ± 7 %. Moreover, purified water, by far being the cheapest available dispersing agent, enables the possibility to replace the so far used organic dispersing agents. This is a major step forward in the direction of the OSWD application for an industrial, large-scale production.

Using the dispersing agent ethylbenzene enables to cover the HOPG surface predominantly with separately occurring 1D QAC chains (refer Fig. 4). The achieved coverage of 24 ± 4 % corresponds to the one of the water based sample. Finally, the dispersing agent dodecane does not catalyse the formation of stable QAC structures. Sometimes or by scanning a large area, tiny sporadically occurring 1D QAC chains can be found, like it's shown in Fig. 13. These structures are not stable, but occur and vanish during the STM scan.

Reviewing these results, we concluded that the dispersing agents deliver activation energy to enable the OSWD process to take place. Furthermore, the dispersing agents direct the OSWD and thereby the structure, size, and coverage of the assembled QAC arrays. Thus, by using appropriate dispersing agents, it is possible to control the self-assembly of the desired 1D or 2D QAC structures in advance. However, the role of the dispersing agent is based on a complex network of diverse factors, including zeta potentials, which needs an extensive discussion and thus will be published elsewhere.

# Journal Name

## ARTICLE

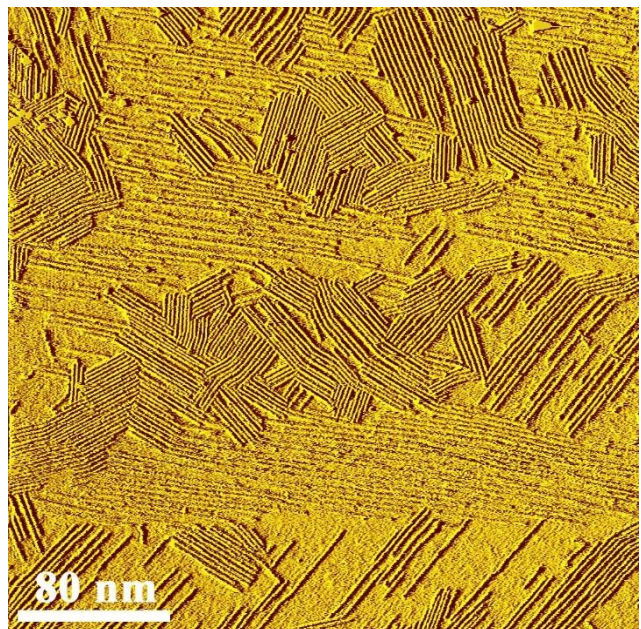

**Fig. 10** STM picture of a HOPG substrate, displaying a surface covering by QAC structures of 67 ± 19 %. Predominant are arrays with a high packing density, the other array configurations hardly occur. Propylene carbonate was used as the catalysing dispersing agent.

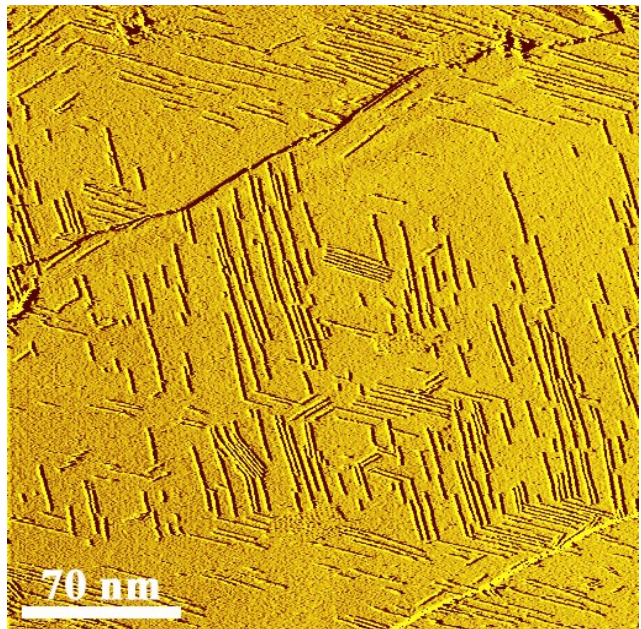

**Fig. 11** STM picture of a HOPG substrate with supramolecular QAC arrays atop, displaying mainly a low packing density. Purified water was used as the dispersing agent. The overall surface coverage is 24 ± 7 %.

**Analysing the stability of the low-dimensional structures and the general applicability of the OSWD**

The results of the performed experiments to analyse the nature of OSWD generated adsorbate structures also enable to estimate their temporal as well as their thermal stability. First of all, the OSWD is performed under ambient conditions and so are the STM measurements used for the sample analysis. This indicates, that the structures are generally resistant to moderate electric fields, as they are applied by the STM, and to humidity. The resistance to water in general is confirmed by the fact, that water can be used as OSWD catalysing dispersing agent. Previously published test results further show, that via the OSWD generated 2D perylenetetracarboxylicdianhydride (PTCDA) arrays do not change their design within a period of at least 3 days.[21] In addition, the QAC samples in this study have been analysed via STM for up to 4 days, the by using propylene carbonate fabricated samples for 7 days, revealing no changes in design or surface coverage. By exploring the thermal stability, it was found, that heating samples up to 160 °C in the presence of different dispersing agents has no observable effect on the overall design of the QAC adsorbate layer. Further tests revealed, that heating up pure γQAC powder on a HOPG substrate to at least 240 °C thermally triggers OSWD processes, leading to an overall surface coverage rate of 83 ± 14 % (Fig. 12). The related samples were again analysed via STM after 36 days, showing still no decomposition of the QAC adsorbates. Summing up, the via the OSWD generated adsorbate structures present themselves to resist temperatures up to at least 240 °C and to have a life span of at least several weeks.

Although the so far presented results were obtained by using a QAC example system, we would like to clarify, that the OSWD is able to process a number of different organic semiconductors such as perinone, flavanthrone, indanthrone, MePTCDI as well as PTCDA and acridone.[21,47]

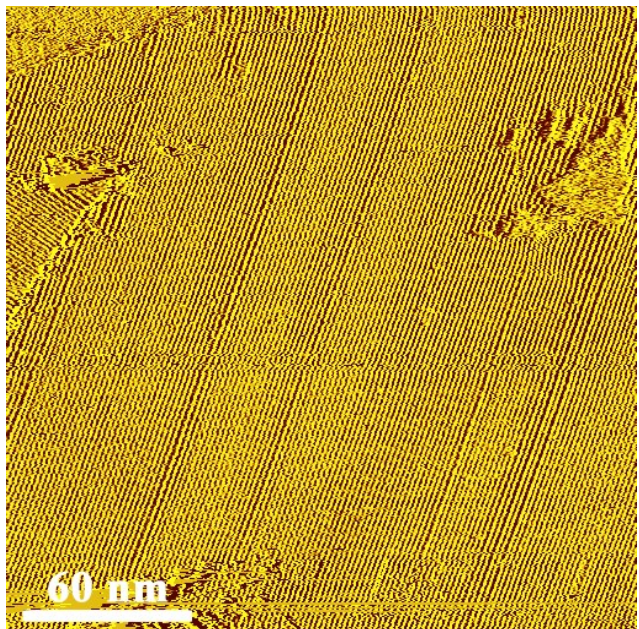

**Fig. 12** QAC adsorbate structures that have been grown by heating up pure γQAC powder on a HOPG to 240 °C. The surface coverage rate is 83 ± 14 %

## Conclusions

Our investigations successfully offered insight into the basic principles and the control capabilities of the OSWD technology. We were able to show that the OSWD enables the assembly of 1D and 2D supramolecular architectures directly from the corresponding 3D crystals. The crystal structures and the lattice parameters of the various QAC array configurations were determined and their dependency on the substrate surface revealed. Further, we were able to verify that the OSWD is applicable to graphite and $MoS_2$ substrates. Additionally, via OSWD assembled supramolecular QAC structures are mechanically stable enough to cover the nanoscale curved surface structure of CNTs.

Working under ambient conditions, the whole OSWD process is driven by the catalysing dispersing agent. Based on our results, we were able to induce the OSWD using low-cost organic dispersing agents, as well as employing purified water. Further, the results indicated that the properties and conditions within the dispersion direct the OSWD and thereby the design and the coverage of the assembled supramolecular structures. Thus, it can be said that the OSWD process can be controlled by choosing the right dispersing agent. In addition, the 1D QAC chains and the 2D arrays form a far-ranging periodic pattern, again dependent on the dispersing agent in use. Such a pattern in turn can be found wherever the dispersion wets the underlying graphite substrate. Hence, the OSWD process can be directed locally and globally, paving a way for a scale up.

Besides, it was found that the generated adsorbate structures are temporally and thermally stable. Further, it can be said that the QAC arrays are chemically and mechanically resistant,[26-32] and can be built up within minutes or even seconds.[20-21] Additionally, previous investigations have shown, that alkyd resin can be used to seal a QAC adsorbate layer, providing a resistant protection layer without damaging the supramolecular structures.[20] Estimating the potentials of the OSWD technology, the above results point towards an industrial application to produce low-cost products using large-scale production technologies, like printed and potentially flexible carbon based electronics,[9] or highly efficient systems to capture carbon dioxide.[45] Especially by using water as dispersing agent, all kinds of non-toxic applications, as in living organism or in the food industry, are imaginable.

## Acknowledgements

The authors would like to thank Ms. Neeti Phatak for proofreading and supporting us in terms of the English language. Further, the Bayerisches Staatsministerium für Umwelt und Verbraucherschutz is gratefully acknowledged for their funding.

## Experimental Section

**Force-field Calculations**

The calculations were performed using the Cerius2 software package employing a Dreiding II force-field,[46] which contains an explicit term for hydrogen bonding. Geometrical constraints were derived from STM measurements and applied to these simulations.

**Scanning electron microscope measurements**

The corresponding measurements have been performed using the scanning electron microscope 440i from Zeiss; the chief scan settings being as: working distance of either 3 or 5 mm and extra-high tension of 20.00 kV. Further, the SE1 detector was used for such measurements.

**STM sample preparation and scan setting**

To prepare a standard scanning tunnelling microscope (STM) sample, using a dispersion with 2 wt % of pigment, γQAC (purchased as Hostaperm Red E5B02 from Clariant) was dispersed in 4 ml of the desired dispersing agent. In this regard, the dispersing agents used were: anisole (from Sigma Aldrich, item no. 10520), ethylbenzene (from Sigma Aldrich, item no. 03080), glycerol (from Alfa Aesar, item no. A1620), propylene carbonate (from Sigma Aldrich, item no. 310328), 8CB (purchased as 4'-n-Octylbiphenyl-4-carbonitrile from Alfa Aesar, item no. 52709-84-9), and purified water. To begin with, a few drops of the dispersion were dispensed on a highly ordered pyrolytic graphite (HOPG, purchased from NT-MDT, item no. GRBS/1.0). In continuative tests, either molybdenum disulphide ($MoS_2$, from Climax Molybdenum Company) or multi-walled carbon nanotubes (CNTs, from ABCR, item no. AB 255407) as alternative substrates were used. The CNTs have been dispersed and ultrasonicated in toluene for 60 minutes. Next, the CNT-dispersion was dropped on a HOPG substrate.

**Journal Name** ARTICLE

After the toluene was completely vaporised (verified in additional STM measurements), the substrate was treated with a γQAC dispersion, in the manner as already described.

If the dispersing agent didn't vaporize after an exposure time of 10 – 20 minutes, the HOPG was dried by a special hotplate, enabling an accurate temperature control and providing a smooth temperature increase (Stuart SD160, temperature accuracy ± 1.0 °C). Once the sample was dried, it was immediately taken off the hotplate. Note, that all dispersing agents vaporized below 160 °C, besides glycerol (235 °C). Heating up pure γQAC powder to 160 °C without a dispersing agent does not induce an OSWD. To create 2D QAC arrays without a dispersing agent was only possible by heating up pure γQAC powder to 240 °C. Thus, we were unable to explore so far, how the high evaporation temperature of glycerol influences the created QAC arrays. Note that the dispersing agent 8CB was not removed, i.e. the corresponding samples haven't been dried, because 8CB does not disturb the STM measurement.

The ready-made STM samples being investigated within days; as per the previous tests, QAC arrays were revealed to not change their structure for a minimum of four weeks, provided they are not influenced via any external forces. For the STM measurements, we used a home-built STM combined with a SPM 100 control system, supplied by RHK Technology. The scans settings were: bias = 1 V, tunnel current = 300 pA, and the line time = 50 ms. Further, the voltage pulses used to improve the scan quality were located in the range between 4.3 and 10 V.

The STM measurements being performed under ambient conditions, a thin layer of dodecane (purchased from Sigma Aldrich, item no. D221104) was generally applied on top of the HOPG surface to increase the measurement quality.[41] Extensive tests revealed, that a mixture of dodecane and γQAC, without another dispersing agent, was unable to generate two-dimensional QAC structures (Fig. 13), and that the dodecane by itself could not form supramolecular assemblies at room temperatures.[42] Besides, to conduct a STM measurement under ambient conditions without dodecane was very challenging, mainly due to the occurrence of a thin contamination layer of adsorbed water on top of the HOPG surface,[38-40] which otherwise was found to be successfully removed using the hydrophobic dodecane.

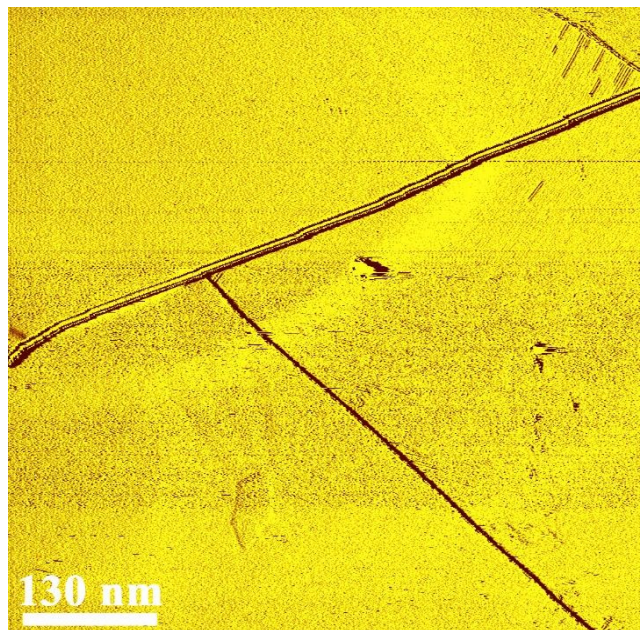

**Fig. 13** STM picture of a HOPG surface, treated with a dispersion of γQAC and the dispersing agent dodecane. No 2D QAC arrays could be found.

**Determining the coverage**

To determine the coverage of the HOPG surface by the QAC arrays within a single STM picture, the software Gwyddion (64bit), version 2.42 was used. For this, initially, the QAC arrays via the tool "Mask Editor" were highlighted, followed by the export of single array dimensions by the tool "Grain distributions", finally accompanied by the Microsoft Excel 2013, version 15.0.4667.1002 calculations to determine the coverage ratio. Further, to investigate the average coverage of a STM sample, we analyzed per sample an area of 0.63 ± 0.24 µm², using a number of STM pictures with high scan resolution and without measurement artefacts; the average coverage rates including the double standard deviations being specified in the current publication. Calculations revealed high doubled standard deviations, indicating the dependence of the coverage by QAC arrays on different positions of a STM sample. Nevertheless, the average coverage rates of different samples treated with different dispersing agents were observed to differ significantly higher in value.

**Fast Fourier Transform**

The Fast Fourier Transform was done using the Fourier processing tool of the software SPIP (Scanning Probe Image Processor, Version 2.3000, distributor: Image Metrology ApS).

# Journal Name

## ARTICLE

**Graphical abstract / table of contents entry**

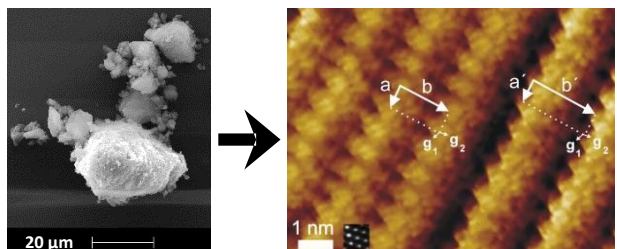

We show that one-dimensional (1D) and two-dimensional (2D) supramolecular crystals can be grown systematically on several substrates under ambient conditions directly from three-dimensional (3D) organic semiconductor crystals.